\documentclass[usenatbib,onecolumn]{mn2e}
\usepackage{epsfig,rotating,natbib,times}

\def\bv{\mathbf{v}}

\def\bB{\mathbf{B}}
\def\br{\mathbf{r}}
\def\bk{\mathrm{\mathbf{k}}}

\def\sma{\sum_a m_a}
\def\smb{\sum_b m_b}
\def\smc{\sum_c m_c}
\def\gwab{\nabla_a W_{ab}}
\def\divB{\nabla\cdot\mathbf{B}}
\newcommand{\pder}[2]{\ensuremath{\frac{\partial #1}{\partial #2}}}

\title[Smoothed Particle Magnetohydrodynamics II]{Smoothed Particle
Magnetohydrodynamics \\ II. Variational principles and
variable smoothing length terms}

\author[Price \& Monaghan]{D.J. Price$^1$, J.J. Monaghan$^2$ \\
$^1$Institute of Astronomy, Madingley Rd, Cambridge, CB3 0HA, UK \\
$^2$School of Mathematical Sciences, Monash University, Clayton 3800, Australia\\
}
\date{Submitted: 6th June 2003; Accepted: 28th October 2003}
\pagerange{\pageref{firstpage}--\pageref{lastpage}} \pubyear{2003}

\begin{document}
\label{firstpage}
\bibliographystyle{/home/dprice/bibtex/mn2e.bst}
\maketitle

\begin{abstract}
In this paper we show how a Lagrangian variational principle can be used
to derive the SPMHD (smoothed particle magnetohydrodynamics) equations for
ideal MHD. We also consider the effect of
a variable smoothing length in the SPH kernels after which we demonstrate by
numerical tests that the consistent treatment of terms relating to the gradient of the
smoothing length in the
SPMHD equations significantly improves the accuracy of the algorithm. Our results
complement those obtained in a companion paper (\citealt{pm03a}, paper I) for
non ideal MHD where artificial dissipative terms were included to handle
shocks.
\end{abstract}


\section{Introduction}
An advantage of deriving numerical algorithms from a variational principle is
that conservation laws can be guaranteed.  Another advantage is that the
algorithms derived from a variational principle  are often more stable than
other algorithms.  For example, in the case of smoothed particle hydrodynamics (SPH, for a
review see \citealt{monaghan92}), the density may be
determined from the continuity equation, and it proves important for stability
to combine the SPH continuity equation with the variational principle to deduce
equations of motion.  We call such a procedure consistent.

  \citet{bl99} have derived consistent SPH equations for fluids even
when non standard forms of the continuity equation are used.  They include the
continuity equation  as a constraint on Lagrangian density variations.  The
resulting equations possess very good stability properties when two fluids with
very different densities, for example air and water, are in contact.  Other,
non consistent, forms of the SPH algorithm, for example with a standard
acceleration equation but non standard continuity equation, exhibit
instabilities. 

In the present paper we show how a Lagrangian variational principle can be used
to derive the SPMHD (smoothed particle magnetohydrodynamics) equations for
ideal MHD.  Variational equations for continuum MHD have been derived by
\citet{newcomb62} for both the Lagrangian and the Eulerian form of the equations
(see also \citealt{henyey82,oppeneer84} and \citealt{field86}).  In the Lagrangian form of the equations Newcomb makes use of flux
conservation to relate changes in the magnetic field to changes in surface
elements.  In the present case, where we consider SPH particles,  it is not
clear how to prescribe such surface elements  in a unique way from the particle
coordinates. Instead we make use of the induction equation in its Lagrangian
form and treat this as a constraint. An alternative, which we do not explore
here, is to begin with plasma physics and prescribe the fields in terms of
currents.  Such an approach would be natural for particle methods (e.g. PIC)
which have been so effective for plasma physics where the electrons would be
treated as one fluid and the ions as another.

The plan of this paper is derive the equations of motion from a standard
Lagrangian for SPH particles with either, or both, the continuity and 
induction equations treated as constraints (\S\ref{sec:sphmom}). We then consider the effect of
variable smoothing length in the SPH kernels (\S\ref{sec:gradh}) after which we demonstrate by 
numerical tests that consistent treatment of the variable smoothing length in the SPH
equations significantly improves the accuracy of SPMHD shocks and of wave propagation
(\S\ref{sec:1Dtests}).  Our results
complement those obtained in a companion paper (\citealt{pm03a}, hereafter paper
I) for non-ideal MHD where artificial dissipative terms were included to handle
shocks.

\section{The Lagrangian}
Variational principles for MHD have been discussed by many authors
(e.g. \citealt{newcomb62,henyey82,oppeneer84,field86}) and the Lagrangian is given by
\begin{equation}
L = \int \left(\frac{1}{2}\rho \mathbf{v}^2 -\rho u-\frac{1}{2\mu_0}B^2\right)
\mathrm{dV},
\end{equation}
which is simply the kinetic minus the potential and magnetic energies. The SPH
Lagrangian is therefore
\begin{equation}
L_{sph} = \smb \left[\frac{1}{2}\mathbf{v}_b^2 - u_b (\rho_b) -\frac{1}{2\mu_0}
\frac{B_b^2}{\rho_b}\right].
\label{eq:spmhdL}
\end{equation}
where we have replaced the integral with a summation and the volume element
$\rho\rm{dV}$ with the mass per SPH particle $m$. Variational principles for SPH
in relativistic and non-relativistic fluid dynamics have been given by
\citet{mp01}.

\section{SPMHD Equations}
\label{sec:sphmom}

\subsection{Equations of motion}

\subsubsection{Standard formulation}
 Ideally we would wish to express all the terms in the Lagrangian
($\ref{eq:spmhdL}$) in terms of the particle co-ordinates, which would
automatically guarantee the conservation of momentum and energy since the
equations of motion result from the Euler-Lagrange equations (e.g. \citealt{mp01}). The density can be written as a function of the particle coordinates
using the usual SPH summation, that is
\begin{equation}
\rho_a = \smb W_{ab},
\label{eq:rhosum}
\end{equation}
where $W_{ab} = W (\br_a - \br_b, h)$ is the SPH interpolation kernel. Taking the
time derivative of this expression, we have the SPH version of the continuity
equation
\begin{equation}
\frac{d\rho_a}{dt} = \smb (\bv_a - \bv_b)\cdot\gwab.
\label{eq:sphcty}
\end{equation}
 The internal energy is regarded as a function of the density, where from the first law of
thermodynamics we have
\begin{equation}
\frac{du}{d\rho} = \frac{P}{\rho^2}.
\label{eq:firstlawthermo}
\end{equation}
The magnetic field is evolved in SPH according to
\begin{equation}
\frac{d\bB_a}{dt} =
\frac{1}{\rho_a}\smb[\bB_a(\bv_{ab}\cdot\gwab)-\bv_{ab}(\bB_a\cdot\gwab)],
\label{eq:Bevolsph1}
\end{equation}
or equivalently
\begin{equation}
\frac{d}{dt}\left(\frac{\bB}{\rho}\right)_a =
-\frac{1}{\rho_a^2}\smb\bv_{ab}(\bB_a\cdot\gwab).
\label{eq:Bevolsph2}
\end{equation}
 (e.g. \citealt{pm85,monaghan92}, paper I). We note that these equations
represent the correct formulation of the induction equation even in the presence of
magnetic monopoles \citep{janhunen00,dellar01}.

However it is not
intuitively obvious how the magnetic field $\bB$ should be related to the particle
co-ordinates, or even that it could be expressed in such a manner (in the SPH
context this would imply an expression for $\bB$ such that taking the time
derivative gives (\ref{eq:Bevolsph1}) or (\ref{eq:Bevolsph2}), analogous to
(\ref{eq:rhosum}) for the density), though it could be done easily for a plasma
with the electrons and ions described by separate sets of SPH particles. We may however proceed by introducing
constraints on $\bB$ in a manner similar to that of \citet{bl99}, that is we require 
\begin{equation}
\delta \int L \mathrm{dt} = \int \delta L \mathrm{dt} = 0,
\label{eq:varprin}
\end{equation}
where we consider variations with respect to a small change in the particle
co-ordinates $\delta \br_a$. We therefore have
\begin{equation}
\delta L = m_a \bv_a\cdot\delta\bv_a - \smb\left[\pder{u_b}{\rho_b}\delta\rho_b +
\frac{1}{2\mu_0}
\left(\frac{B_b}{\rho_b}\right)^2\delta\rho_b - \frac{1}{\mu_0}
\bB_b\cdot\delta\left(\frac{\bB_b}{\rho_b}\right)\right].
\label{eq:deltaL}
\end{equation}
The Lagrangian variations in density and magnetic field are given by 
\begin{eqnarray}
\delta\rho_b & = & \smc\left(\delta \br_b - \delta
\br_c\right)\cdot \nabla_b W_{bc}
\label{eq:deltarho} \\
\delta\left(\frac{\bB_b}{\rho_b}\right) & = & \smc (\delta
\br_b - \delta \br_c) \frac{\bB_b}{\rho_b^2} \cdot\nabla_b W_{bc} 
\label{eq:deltaBrho}
\end{eqnarray}
where we have used (\ref{eq:sphcty}) and (\ref{eq:Bevolsph2}) respectively (note
that we also recover the following results if we use (\ref{eq:Bevolsph1}) instead of
(\ref{eq:Bevolsph2})). Using (\ref{eq:deltarho}), (\ref{eq:deltaBrho}) and (\ref{eq:firstlawthermo}) in (\ref{eq:deltaL}) and rearranging, we find
\begin{eqnarray}
\frac{\delta L}{\delta \br_a} & = & -\smb\left[\frac{P_b}{\rho_b^2}\smc \nabla_b W_{bc} (\delta_{ba} -
	 \delta_{ca}) \right] \nonumber \\
         & - & \smb\left[\frac{1}{2\mu_0}\left(\frac{\bB_b}{\rho_b}\right)^2
	 \nabla_b W_{bc} (\delta_{ba} - \delta_{ca})\right]\nonumber \\
	 & + & \smb\left[ \frac{1}{\mu_0}\frac{\bB_b}{\rho_b^2}\smc \frac{\bB_b}{\rho_b}\cdot \nabla_b W_{bc}(\delta_{ba} -
	 \delta_{ca})\right],
\end{eqnarray}
where $\delta_{ab}$ refers to the Kronecker delta. Putting this back into
(\ref{eq:varprin}), integrating the velocity term by
parts and simplifying (using $\gwab = -\nabla_b W_{ba}$), we obtain
\begin{eqnarray}
\int\left\{-m_a \frac{d\bv_a}{dt} \right. & - & \smb \left(\frac{P_a}{\rho_a^2} + \frac{P_b}{\rho_b^2} \right) \nabla_a W_{ab} \nonumber \\
& - & \smb \frac{1}{2\mu_0}\left(\frac{B_a^2}{\rho_a^2} + \frac{B_b^2}{\rho_b^2}\right)\gwab \nonumber \\
& + & \left. \smb \frac{1}{\mu_0}\left[\frac{\bB_a}{\rho_a^2} (\bB_a\cdot\gwab) +
\frac{\bB_b}{\rho_b^2} (\bB_b\cdot\gwab) \right]\right\}\delta\br_a \mathrm{dt} = 0.
\end{eqnarray}
The SPH equations of motion are therefore given by
\begin{equation}
\frac{dv^i_a}{dt} = \smb\left[\left(\frac{S^{ij}}{\rho^2}\right)_a +
\left(\frac{S^{ij}}{\rho^2}\right)_b\right]\nabla_a^j W_{ab},
\label{eq:tensor}
\end{equation}
where the stress tensor $S^{ij}$ is defined as
\begin{equation}
S^{ij} \equiv -P\delta^{ij} + \frac{1}{\mu_0}\left(B^i B^j -
\frac{1}{2}B^2\delta^{ij}\right). \label{eq:mij}
\end{equation}

This form of the magnetic force term conserves linear momentum
exactly (angular momentum is discussed in \S\ref{sec:spmhdangmom}) but was shown by \citet{pm85} to be unstable in certain
regimes (low magnetic $\beta$). We resolve this instability by adding a short range repulsive force to prevent 
particles from clumping \cite{monaghan00}, the implementation of which is described in
paper I. We note that the
conservative form of the momentum equation was derived using a non-conservative induction
equation, which agrees with the treatment of magnetic monopoles suggested by \citet{janhunen00} and
\citet{dellar01}.

\subsubsection{Alternative formulation}
\label{sec:altspmhd}
 Consistent sets of SPMHD equations may also be derived using alternative forms of
the continuity and induction equations. We give the example below since alternative
forms of the pressure terms in the momentum equation are often explored in the
context of SPH, without alteration of the other equations to make the formalisms
self-consistent. We expect that a lack of consistency in the discrete equations
will inevitably lead to loss of accuracy in the resulting algorithm.
For example, using the continuity equation
\begin{equation}
\frac{d\rho_a}{dt} = \rho_a \smb \frac{\bv_{ab}}{\rho_b} \cdot \gwab,
\label{eq:sphctyalt}
\end{equation}
and the induction equation
\begin{equation}
\frac{d}{dt}\left(\frac{\bB}{\rho}\right)_a =
-\frac{1}{\rho_a}\smb\frac{\bv_{ab}}{\rho_b}(\bB_a\cdot\gwab).
\label{eq:Bevolsphalt}
\end{equation}
results in the momentum equation
\begin{equation}
\frac{dv^i_a}{dt} = \smb \left[\frac{S^{ij}_a +
S^{ij}_b}{\rho_a\rho_b}\right]\nabla^j_a W_{ab}.
\end{equation}
This form of the SPMHD equations also conserves linear momentum exactly
and in the hydrodynamic case has been found to give better performance in
situations where there are large jumps in density (for example at a water-air
interface). The consistent form of the energy equations are given in
\S\ref{sec:altenergy}.

\subsection{Energy equation}
\label{sec:spmhdener}

\subsubsection{Internal energy}
The internal energy
equation follows from the use of the first law of thermodynamics, that is
\begin{equation}
\frac{du_a}{dt} = \frac{P_a}{\rho_a^2}\frac{d\rho_a}{dt}.
\end{equation}
Using the standard continuity equation (\ref{eq:sphcty}) therefore gives
\begin{equation}
\frac{du_a}{dt} = \frac{P_a}{\rho_a^2} \smb \bv_{ab}\cdot\gwab.
\label{eq:sphutherm}
\end{equation}

\subsubsection{Total energy}
 The Hamiltonian is given by
\begin{equation}
H = \sum_a \bv_a \cdot \pder{L}{\bv_a} - L.
\label{eq:H}
\end{equation}
which represents the conserved total energy of the SPH particles since the
Lagrangian does not explicitly depend on the time co-ordinate. Using (\ref{eq:spmhdL}) we have
\begin{eqnarray}
H = E = \sma \left(\frac{1}{2}v_a^2 + u_a + \frac{1}{2}\frac{B_a^2}{\rho_a}\right).
\end{eqnarray}
Taking the (comoving) time derivative, we have
\begin{equation}
\frac{dE}{dt} = \sma \left[\bv_a\cdot\frac{d\bv_a}{dt} +
\frac{du_a}{d\rho_a}\frac{d\rho_a}{dt} +
\frac{1}{2}\frac{B_a^2}{\rho_a^2}\frac{d\rho_a}{dt} +
\bB_a\cdot\frac{d}{dt}\left(\frac{\bB_a}{\rho_a}\right)\right],
\end{equation}
where the first term is specified by use of the momentum equation
(\ref{eq:tensor}), the second term
using the first law of thermodynamics (\ref{eq:firstlawthermo}) and the
continuity equation (\ref{eq:sphcty}), the third term by
the continuity equation (\ref{eq:sphcty}) and the fourth term by the induction
equation (\ref{eq:Bevolsph2}). Using these equations and simplifying we find
\begin{equation}
\frac{dE}{dt} = \sma \smb \left[\left(\frac{S^{ij}}{\rho^2}\right)_a v_b^i
+ \left(\frac{S^{ij}}{\rho^2}\right)_b v_a^i\right] \nabla^j_a W_{ab},
\end{equation}
such that the total energy per particle is evolved according to
\begin{equation}
\frac{d\hat{\epsilon}_a}{dt} = \smb \left[\left(\frac{S^{ij}}{\rho^2}\right)_a v_b^i
+ \left(\frac{S^{ij}}{\rho^2}\right)_b v_a^i\right] \nabla^j_a W_{ab},
\label{eq:energymhd}
\end{equation}
where
\begin{equation}
\hat{\epsilon}_a = \frac{1}{2}v_a^2 + u_a + \frac{1}{2}\frac{B_a^2}{\rho_a}
\end{equation}
is the energy per unit mass.

\subsubsection{Alternative formulation}
\label{sec:altenergy}
 For the alternative formulation given in \S\ref{sec:altspmhd} the internal energy
equation is given by
\begin{equation}
\frac{du_a}{dt} = \frac{P_a}{\rho_a} \smb \frac{\bv_{ab}}{\rho_b} \cdot \gwab,
\end{equation}
and the total energy equation by
\begin{equation}
\frac{d\hat{\epsilon}_a}{dt} = \smb \left[ \frac{S^{ij}_a v^i_b + S^{ij}_b
v^i_a}{\rho_a\rho_b} \right] \nabla^j_a W_{ab}.
\end{equation}

\section{Variable smoothing length terms}
\label{sec:gradh}
 The smoothing length $h$ determines the radius of interaction for each SPH
particle. Early SPH simulations used a fixed smoothing length for all particles,
however allowing each particle to have its own associated smoothing length which
varies according to local conditions increases the spatial resolution
substantially \citep{hk89,benz90}. The usual rule is to take
\begin{equation}
h_a \propto \left(\frac{1}{\rho_a} \right)^{(1/\nu)},
\label{eq:hrho}
\end{equation}
where $\nu$ is the number of spatial dimensions, although others are possible
\citep{monaghan00}. Implementing this rule
self-consistently is more complicated in SPH since the density $\rho_a$ is itself a
function of the smoothing length $h_a$ via the relation (\ref{eq:rhosum}). The
usual rule is to take the time derivative of (\ref{eq:hrho}), giving (e.g.
\citealt{benz90})
\begin{equation}
\frac{dh_a}{dt} = -\frac{h_a}{\nu\rho_a}\frac{d\rho}{dt},
\label{eq:hevol}
\end{equation}
which can then be evolved alongside the other particle quantities.

This rule works well for most practical purposes, and maintains the relation (\ref{eq:hrho})
particularly well when the density is updated using the continuity equation
(\ref{eq:sphcty}). However, it has been known for
some time that, in order to be fully self-consistent, extra terms involving the
derivative of $h$ should be included in the momentum and energy equations (e.g.
\citealt{nelson94,np94,sea96}). Attempts to do this were, however, complicated to
implement \citep{np94} and therefore not generally adopted by the SPH
community. Recently \citet{sh02} have shown that the so-called $\nabla h$
terms can be self-consistently included in the equations of motion and energy
using a variational approach. \citet{sh02} included the variation of the
smoothing length in their variational principle by use of Lagrange multipliers,
however, in the context of the discussion given in \S\ref{sec:sphmom} we note
that by expressing the smoothing length as a function of $\rho$ we can therefore
specify $h$ as a function of the particle co-ordinates \citep{monaghan02}. That
is we have $h = h(\rho)$ where $\rho$ is given by
\begin{equation}
\rho_a = \smb W(\br_{ab}, h_a).
\label{eq:rhosumgradh}
\end{equation}
Taking the time derivative, we obtain
\begin{equation}
\frac{d\rho_a}{dt} = \frac{1}{\Omega_a} \smb \bv_{ab}\cdot\nabla_a W_{ab}(h_a),
\label{eq:sphctygradh}
\end{equation}
where
\begin{equation}
\Omega_a = \left[1 + \pder{h_a}{\rho_a}\smc
\pder{W_{ab}(h_a)}{h_a}\right]^{-1}.
\label{eq:omegaa}
\end{equation}

The equations of motion in the hydrodynamic case may then be found using the Euler-Lagrange equations
and will automatically conserve linear and angular momentum. The
resulting equations are given by \citep{sh02,monaghan02}
\begin{equation}
\frac{d\bv_a}{dt} = - \smb \left[ \frac{P_a}{\Omega_a\rho_a^2}\nabla_a
W_{ab}(h_a) + \frac{P_b}{\Omega_b\rho_b^2}\nabla_a W_{ab}(h_b)\right].
\label{eq:sphmomgradh}
\end{equation}

Calculation of the quantities $\Omega$ involve a summation over the particles and can be
computed alongside the density summation (\ref{eq:rhosumgradh}). To be fully self-consistent
(\ref{eq:rhosumgradh}) should be solved iteratively to determine both $h$
and $\rho$ self-consistently. In practice this means that an extra pass over the density summation only occurs
when the density changes significantly between timesteps. \citet{sh02} also suggest using the
continuity equation (\ref{eq:sphctygradh}) to obtain a better starting approximation for $\rho$ and consequently 
$h$. We perform simple fixed point iterations of the density, using a predicted smoothing
length from (\ref{eq:hevol}). Having calculated the density by summation, we then
compute a new value of smoothing length $h_{new}$ using (\ref{eq:hrho}). The convergence of each
particle is then determined according to the criterion
\begin{equation}
\frac{\vert h_{new} - h\vert}{h} < 1.0 \times 10^{-2}.
\end{equation}
We then iterate until all particles are converged, although for efficiency we do not allow the scheme
to continue iterating on the same particle(s). Note that a particle's smoothing
length is only set equal to $h_{new}$ if the density is to be recalculated (this is to
ensure that the same smoothing length that was used to calculate the density is used to
compute the terms in the other SPMHD equations). The calculated gradient terms (\ref{eq:omegaa}) may also be used to implement an iteration scheme such as the
Newton-Raphson method which converges faster than our simple fixed point
iteration. We also note that in principle only the density on particles which have not converged
need to be recomputed, since each particle's density is independent of the smoothing
length of neighbouring particles. These considerations will be discussed further in the multidimensional
context since the cost of iteration is of greater importance in this case.

 Since we cannot explicitly write the Lagrangian (\ref{eq:spmhdL}) as a function
of the particle co-ordinates, we cannot explicitly derive the SPMHD equations
incorporating a variable smoothing length. We may, however deduce the form of the terms
which should be included by consistency arguments. We start with the SPH induction equation
in the form
\begin{equation}
\frac{d}{dt}\left(\frac{\bB}{\rho}\right)_a =
-\frac{1}{\rho_a^2}\smb\bv_{ab}(\bB_a\cdot\gwab).
\label{eq:brhonormal}
\end{equation}
Expanding the left hand side, we have
\begin{equation}
\frac{d\bB_a}{dt} = -\frac{1}{\rho_a}\smb\bv_{ab}(\bB_a\cdot\gwab) - \frac{\bB_a}{\rho_a}\frac{d\rho_a}{dt}.
\label{eq:Bevolgradh}
\end{equation}
If the smoothing length is a given function of the density, then the SPH
continuity equation is given by (\ref{eq:sphctygradh}) and (\ref{eq:Bevolgradh}) becomes
\begin{equation}
\frac{d\bB_a}{dt} = -\frac{1}{\rho_a}\smb \left\{\bv_{ab}(\bB_a\cdot\gwab) -
\frac{1}{\Omega_a} \bB_a [\bv_{ab}\cdot \nabla_a W_{ab}(h_a)]\right\}.
\end{equation}
However in one dimension these terms must cancel to give $B_x =$ const, and thus
we deduce that the correct form of the induction equation is therefore
\begin{equation}
\frac{d\bB_a}{dt} = -\frac{1}{\Omega_a\rho_a}\smb \left\{\bv_{ab}
\left[\bB_a\cdot\nabla_a W_{ab}(h_a)\right] -
\bB_a \left[\bv_{ab}\cdot\nabla_a W_{ab}(h_a)\right]\right\},
\label{eq:Bgradh}
\end{equation}
or in the form (\ref{eq:brhonormal}) we would have
\begin{equation}
\frac{d}{dt}\left(\frac{\bB}{\rho}\right)_a =
-\frac{1}{\Omega_a\rho_a^2}\smb\bv_{ab}[\bB_a\cdot\nabla_a W_{ab}(h_a)].
\label{eq:Brhogradh}
\end{equation}
 Using (\ref{eq:Bgradh}) or
(\ref{eq:Brhogradh}) and (\ref{eq:sphctygradh}) as constraints we may then derive the equations of motion
using the variational principle described in \S\ref{sec:sphmom} to give
\begin{equation}
\frac{d\bv_a}{dt} = \smb \left[ \left(\frac{S^{ij}}{\Omega\rho^2}\right)_a
\nabla^j_a W_{ab}(h_a) + \left(\frac{S^{ij}}{\Omega\rho^2}\right)_b
\nabla^j_a W_{ab}(h_b) \right].
\label{eq:spmhdmomgradh}
\end{equation}
The total energy equation is given by
\begin{equation}
\label{eq:energradh}
\frac{d\hat{\epsilon}_a}{dt} = \smb \left[ \left(\frac{S^{ij}}{\Omega\rho^2}\right)_a v^i_b
\nabla^j_a W_{ab}(h_a) + \left(\frac{S^{ij}}{\Omega\rho^2}\right)_b v^i_a
\nabla^j_a W_{ab}(h_b) \right],
\end{equation}
whilst the internal energy equation is found using the first law of
thermodynamics and (\ref{eq:sphctygradh}), that is
\begin{equation}
\frac{du_a}{dt} = \frac{P_a}{\Omega_a \rho_a^2} \smb \bv_{ab}\cdot\nabla_a W_{ab}(h_a)
\label{eq:uthermgradh}
\end{equation}

 We show in \S\ref{sec:mwav} that including the correction terms for a variable
smoothing length in this manner significantly improves the numerical wave speed
in the propagation of MHD waves and enables the shock tube problems considered in
paper I to be computed with no smoothing of the initial conditions.

\section{Momentum conservation}
\label{sec:spmhdangmom}
The equations of motion conserve linear momentum exactly.  However, angular
momentum is not conserved exactly because the stress force between  a pair of
particles is not along the line joining them.  Returning to
(\ref{eq:spmhdmomgradh}), and
considering motion in 2 dimensions $x$ and $y$, the change in angular momentum
of the system is given by
\begin{equation}
\frac{d}{dt} \sum_a ( {\bf r}_a \times {\bf v}_a)^z  =   \sum_a \sum_b m_a m_b \left ( \left [\bar{\sigma}_{ab}^{xx} -\bar{\sigma}_{ab}  ^{yy} \right ] y_{ab}  x_{ab} + \bar{\sigma}_{ab}^{xy} [y^2_{ab} - x^2_{ab} ] \right )  F_{ab},
\label{eq:angmom}
\end{equation}
where $y_{ab} = y_a - y_b$,  $x_{ab} = x_a - x_b$,  $\sigma^{ij} = S^{ij}/
\Omega \rho^2$ and $\tilde\sigma^{ij}_{ab} = \sigma_a^{ij} + \sigma_b^{ij} $. 
We have replaced  $\nabla W_{ab}$ by ${\bf r}_{ab} F_{ab}$.  It can be seen
from (\ref{eq:angmom}) that if the stress is isotropic, and proportional to the identity
tensor, as is the case for isotropic fluids, the rate of change of angular
momentum vanishes.  If, however, the stress is not proportional to the identity
tensor then the total angular momentum of the system will change.  It can be
shown that when the summations can be converted to integrals the angular
momentum is conserved exactly. 

The same problem arises in the case of elastic stresses where the problem is
exacerbated by the fact that the particles near the edge of a solid have
densities similar to the interior and the particles do not have neighbours
exterior to the solid. In this case  the conservation of angular momentum is
significantly in error. \citet{bl99} showed, however, that angular
momentum could be conserved by altering the gradient of the kernel to a matrix
operator.  The astrophysical problem could be solved in the same way but we
expect the astrophysical conservation to be very much better without changing
the kernel,  because edges are associated with low density and correspondingly
low angular momentum.  

\section{Numerical tests}
\label{sec:1Dtests}
 We demonstrate the usefulness of the variable smoothing length terms in the
MHD case by the simulation of MHD waves and the \citet{bw88} shock tube problem.
We find that with the variable smoothing length terms included it is better to use 
(\ref{eq:Brhogradh}) to update the magnetic field rather than (\ref{eq:Bgradh}) since
we find that using (\ref{eq:Bgradh}) can lead to negative thermal energies in the shock tube problem. 
The results shown use the total energy equation (as in paper I) although the
similar results are obtained when the thermal energy is integrated.

\subsection{MHD waves}
\label{sec:mwav}
 The equations of magnetohydrodynamics admit three `families' of waves, the
so called slow, alfven and fast waves (appendix \ref{sec:appendixA}). 
 The tests presented here are taken from \citet{dw98}. We consider travelling slow
and fast MHD waves propagating in a 1D domain, where the velocity and magnetic field 
are allowed to vary in three dimensions. We use $\gamma = 5/3$ for the problems
considered here. The perturbation in density is
applied by perturbing the particles from an initially uniform setup (since we
use equal mass particles). Details
of this perturbation are given in appendix \ref{sec:appendixB} and the amplitudes for the
other quantities are derived in
appendix \ref{sec:appendixA}. We leave the artificial dissipation on for this
problem, with the \citet{mm97} switch implemented using $K_{min} = 0.05$ (see
paper I for details of this implementation). This is to
demonstrate that the switch is effective in turning off the
artificial dissipation in the absence of shocks. The variable smoothing length terms
(\S\ref{sec:gradh})
do not affect the wave profiles but inclusion of these terms gives
very accurate numerical wave speeds.

\begin{figure*}[t]
\begin{center}
\epsfig{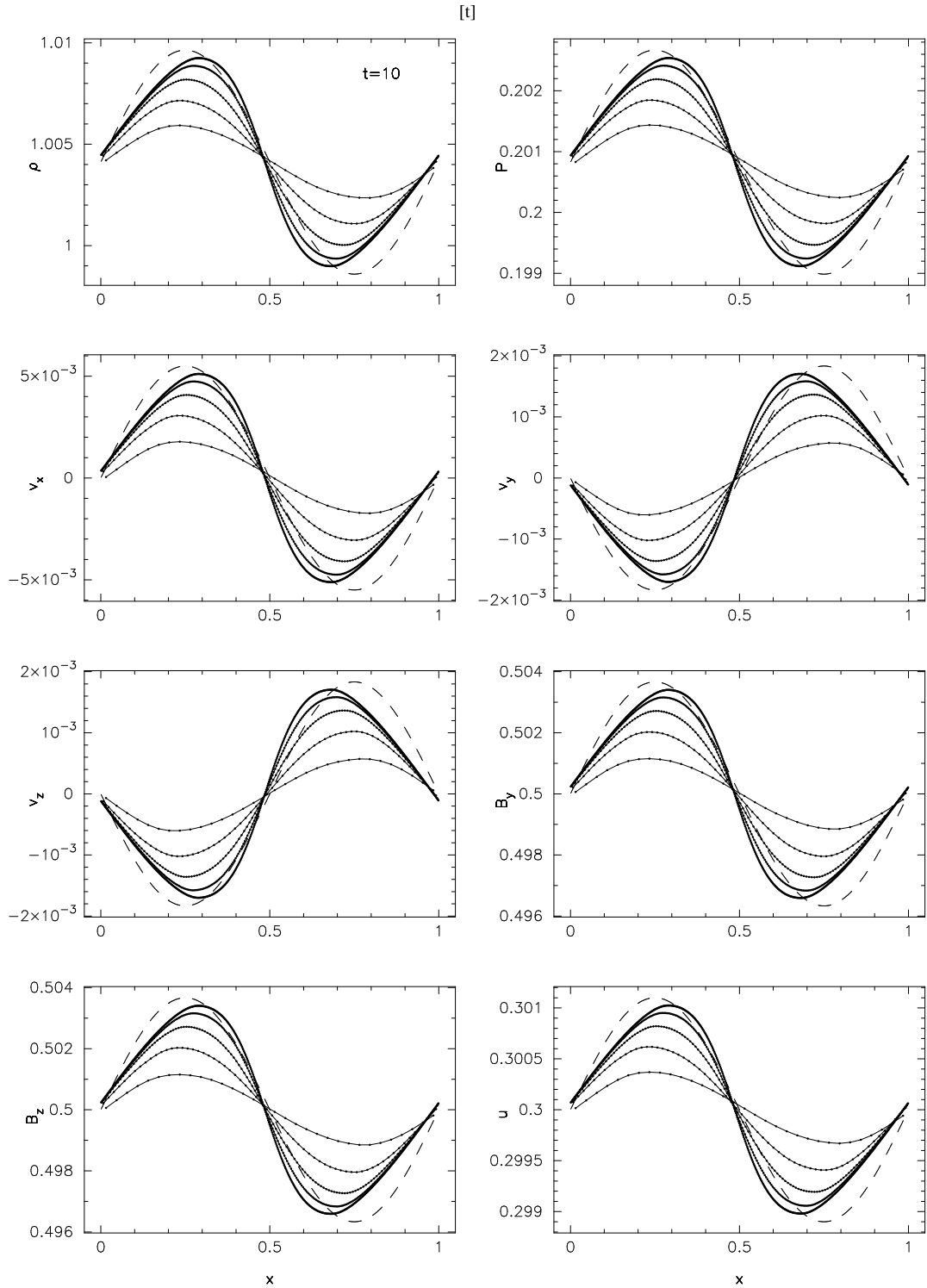}
\caption{Results for the 1D travelling fast wave problem. Initial conditions are indicated by the dashed line. Results are presented after 10 periods for simulations with
32, 64, 128, 256 and 512 particles. The fast wave speed in the gas is very close to
unity which is accurately reproduced by the SPMHD solution (ie. the numerical solution is
in phase with the initial conditions). The artificial dissipation is turned on but uses
the switch of \citet{mm97} which dramatically reduces its effects away from
shocks. The wave is steepened slightly by nonlinear effects.}
\label{fig:fastwave}
\end{center}
\end{figure*}

The fast wave is shown in Figure \ref{fig:fastwave},
with the dashed line giving the initial conditions. The initial amplitude is
0.55\% as in \citet{dw98}. Results are shown at 
t=10 for five different simulations using 32, 64, 128, 256 and 512 particles in the
x-direction. The properties of the gas are set
such that the fast wave speed is very close to unity, meaning that the solution at
$t=10$ should be in phase with the initial conditions. Figure \ref{fig:fastwave}
demonstrates that this is accurately reproduced by the SPMHD algorithm. The effects of the small amount of
dissipation present can be seen in the amount of damping present in the solutions. The small amount of steepening observed
wave profiles is due to nonlinear effects and agrees with the results presented by
\citet{dw98}.

\begin{figure*}[t]
\begin{center}
\epsfig{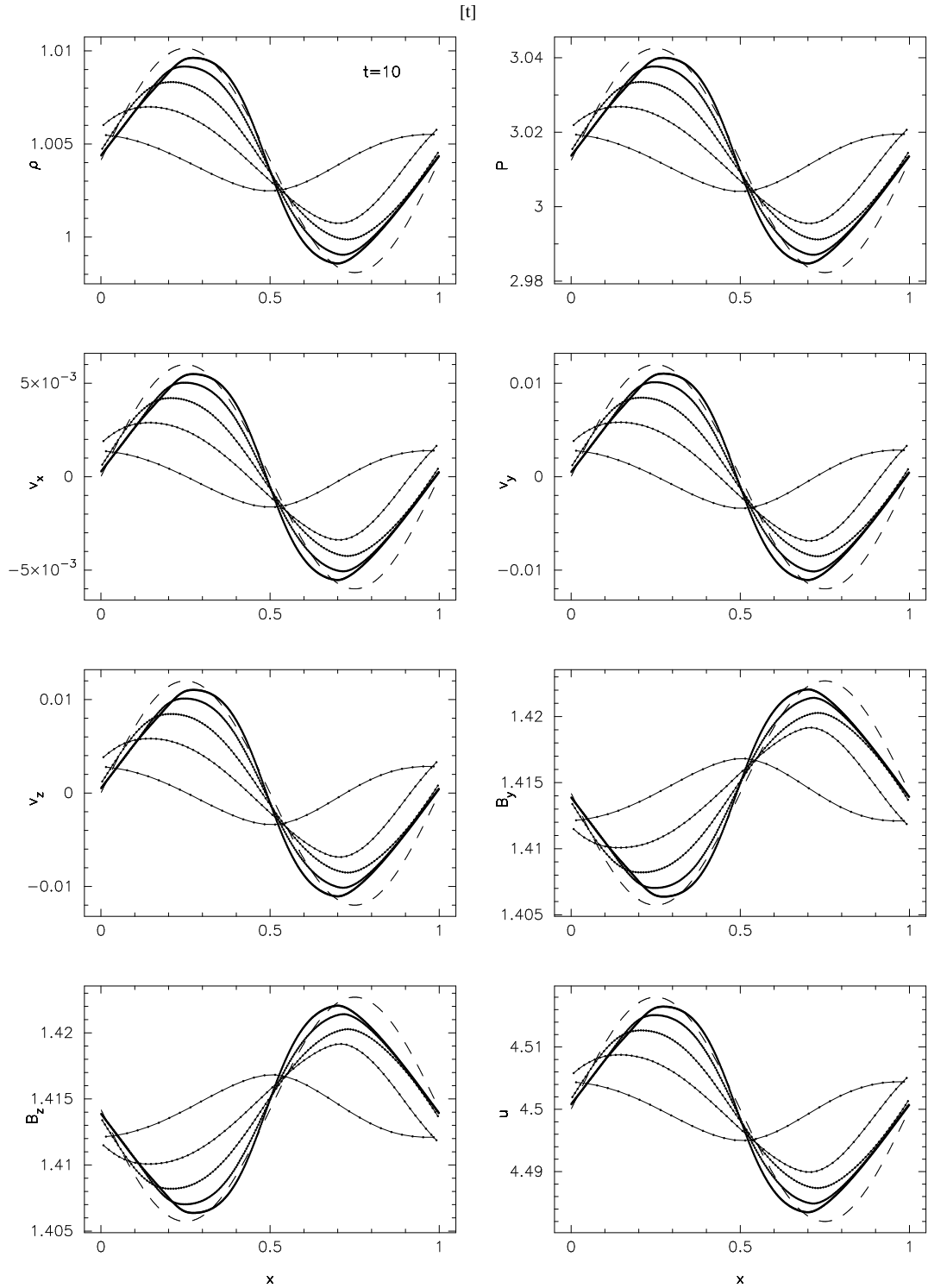}
\caption{Results for the 1D travelling slow wave problem. Initial conditions are
indicated by the dashed line and results are presented after 10 periods for simulations with
32, 64, 128, 256 and 512 particles. The slow wave speed in the gas is very close to
unity, such that the numerical solution at $t=10$ should be in phase with the initial
conditions. This is well represented by the SPMHD solution for the higher resolution
runs. The artificial dissipation is turned on but uses
the switch of \citet{mm97} which dramatically reduces its effects away from
shocks. The wave is steepened slightly by nonlinear effects.}
\label{fig:slowwave}
\end{center}
\end{figure*}

 The slow MHD wave is shown in Figure \ref{fig:slowwave}, again with the
dashed line giving the initial conditions. The perturbation amplitude is 0.6\% as in
\citet{dw98}. Results are again shown at $t=10$ at
resolutions of 32, 64, 128, 256 and 512 particles.In this case the properties of the gas
being set such that the slow wave speed in the medium is very close to unity, again meaning
that the solution at $t=10$ should be in phase with the initial conditions. We see in
Figure \ref{fig:slowwave} that this is reproduced by the SPMHD solution for the higher
resolution runs. The artificial dissipation 
is again turned on using the switch and a minimum of $K_{min} = 0.05$. The wave is slightly
overdamped in this case since we construct the artificial dissipation using the
fastest wave speed (c.f. paper I) which in this case is approximately three times
the wave propagation speed. 

\subsection{Shock tube}
\begin{figure*}[t]
\begin{center}
\epsfig{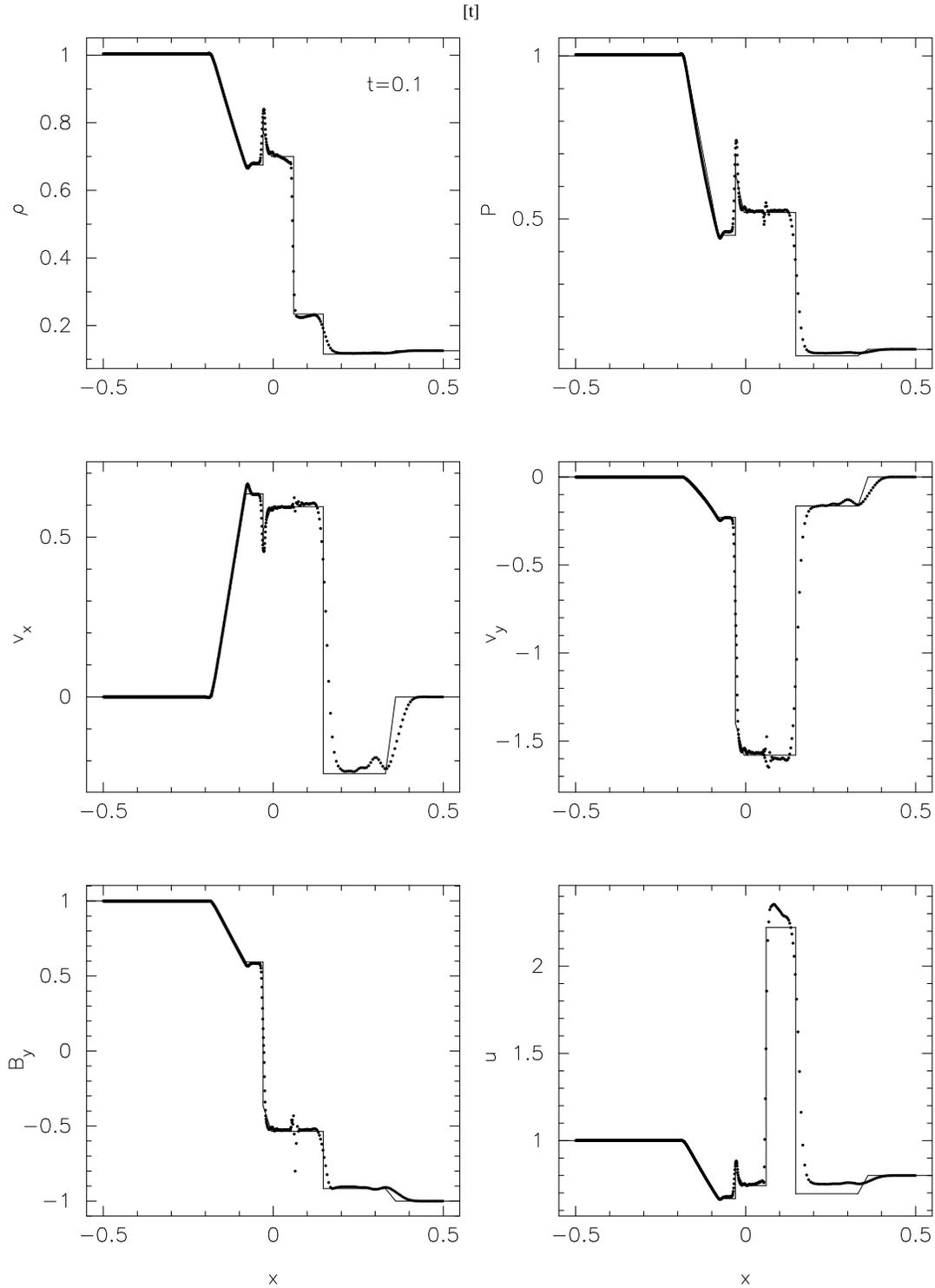}
\caption{Results of the \citet{bw88} shock tube test with no smoothing of the
initial conditions. Initial conditions to the
left of the origin are given by $(\rho,P,v_x,v_y,B_y) = [1,1,0,0,1]$ and to the
right by $(\rho,P,v_x,v_y,B_y)=[0.125,0.1,0,0,-1]$ with
$B_x = 0.75$ and $\gamma=2.0$. Profiles of density, pressure, $v_x$, $v_y$,
transverse magnetic field and thermal energy are shown at time $t=0.1$ and may be
compared with the numerical solution from \citet{balsara98} given by the solid line. In this case the density
summation, total energy equation and the induction equation using $\bB/\rho$
have been used, incorporating the variable smoothing length terms.}
\label{fig:briowu}
\end{center}
\end{figure*}

 As an additional example of the advantages of the consistent smoothing length
evolution and the variable smoothing length terms we recalculate the \citet{bw88}
shock tube test from paper I. In this case however we apply no smoothing
whatsoever to the initial conditions and calculate the solution using the
density summation (\ref{eq:rhosumgradh}), the total energy equation
(\ref{eq:energradh}) and the induction equation
(\ref{eq:Brhogradh}). As in paper I we set up the problem using approximately
800 equal mass particles in the domain $x =
[-0.5, 0.5]$. Conditions to the left of the shock are given by $(\rho,P,v_x,v_y,B_y) = [1,1,0,0,1]$ and
to the right by $(\rho,P,v_x,v_y,B_y)=[0.125,0.1,0,0,-1]$ with
$B_x = 0.75$ and $\gamma=2.0$. The results are shown in Figure \ref{fig:briowu}
at time $t=0.1$ and may be compared with the numerical solution from \citet{balsara98}
given by the solid line. The results may also be compared with Figure 2 in paper I. The
non-smoothed initial conditions result in a small starting error at the contact
discontinuity and a small overshoot at the end of the rarefaction wave, however
the compound wave in particular is significantly less spread out than in the
results given in paper I. The consistent update of the smoothing length
discussed in \S\ref{sec:gradh} results in some extra iterations of the density
(for most of the simulation two passes over the density summation are used).

\begin{figure*}[t]
\begin{center}
\epsfig{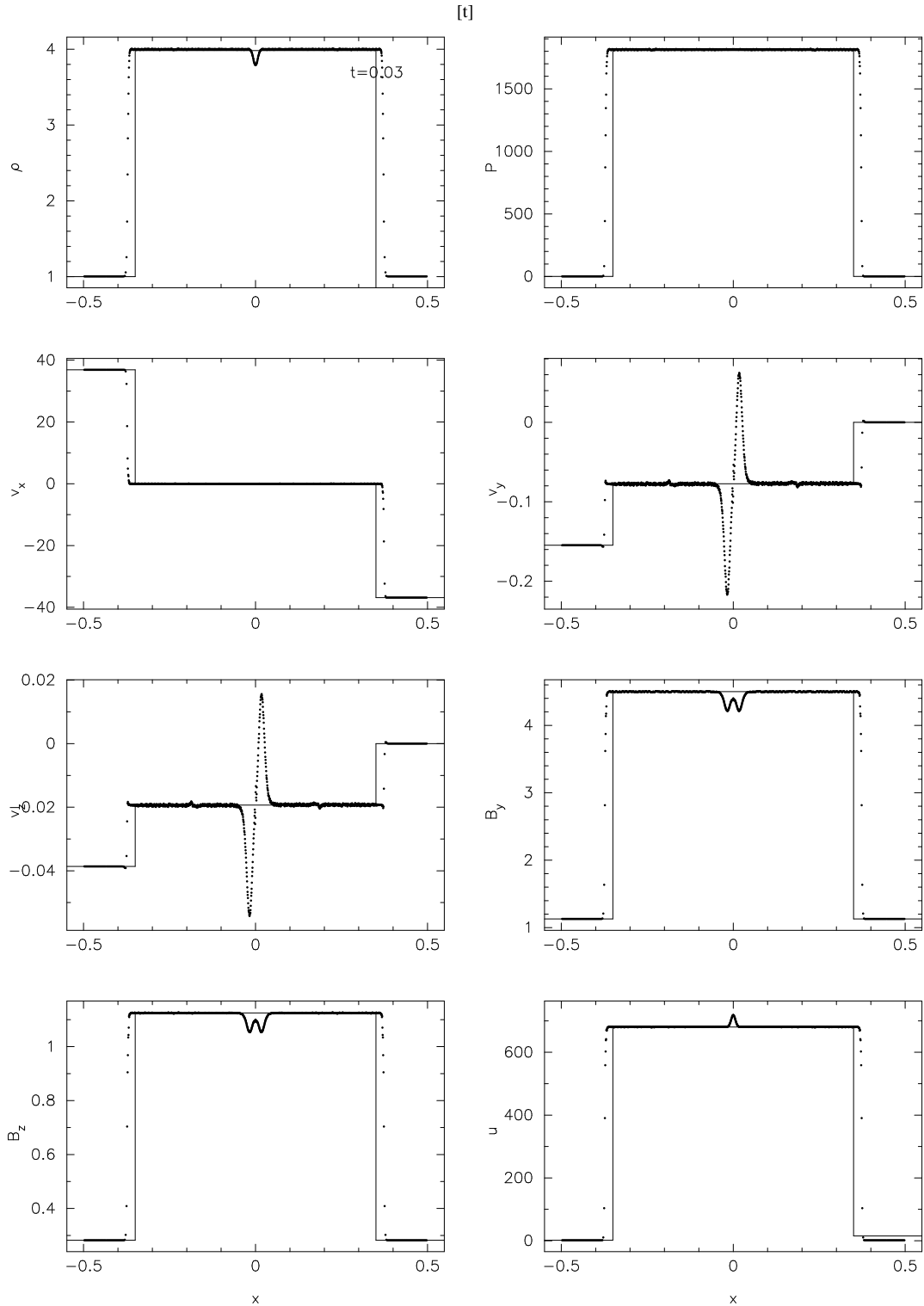}
\caption{Results of the MHD shock tube test with initial conditions to the left of the
shock given by $(\rho,P,v_x,v_y,v_z,B_y,B_z) =
[1,1,36.87,-0.155,-0.0386,4/(4\pi)^{1/2},1/(4\pi)^{1/2}]$ and to the right by
$(\rho,P,v_x,v_y,v_z,B_y,B_z) = [1,1,-36.87,0,0,4/(4\pi)^{1/2},1/(4\pi)^{1/2}]$ with
$B_x = 4.0/(4\pi)^{1/2}$  and $\gamma = 5/3$. Results
are shown at time $t=0.03$ compare extremely well with the exact solution
given by \citet{dw94} (solid lines). The overshoots in density,
pressure and magnetic field observed in paper I are no longer present due to our
self-consistent inclusion of terms relating to the gradient of the smoothing length.}
\label{fig:balsara4}
\end{center}
\end{figure*}

 As a final example we also recompute the shock tube test shown in Figure 7 of paper
I. The initial conditions to the left of the shock are given by
$(\rho,P,v_x,v_y,v_z,B_y,B_z) =
[1,1,36.87,-0.155,-0.0386,4/(4\pi)^{1/2},1/(4\pi)^{1/2}]$ and to the right by
$(\rho,P,v_x,v_y,v_z,B_y,B_z) = [1,1,-36.87,0,0,4/(4\pi)^{1/2},1/(4\pi)^{1/2}]$ with
$B_x = 4.0/(4\pi)^{1/2}$ and $\gamma = 5/3$, resulting in two extremely strong fast
shocks which propagate away from the origin. The resolution varies from 400 to 1286
particles throughout the simulation due to the inflow boundary conditions. Results are shown in Figure
\ref{fig:balsara4} at time $t=0.03$ and compare extremely well with the exact solution
given by \citet{dw94} (solid lines). In paper I the post-shock density and transverse
magnetic field components were observed to overshoot the exact solution. In Figure
\ref{fig:balsara4} we observe that these effects are no longer present when the
variable smoothing length terms are self-consistently accounted for.

\section{Summary}
 In summary, we have shown that
 
\begin{enumerate}
\item  The equations of motion and energy for SPMHD can be derived from a variational
principle using the continuity and induction equations as constraints. This demonstrates
that the equation set is consistent and the resulting equations conserve
linear momentum and energy exactly. In the MHD case this also demonstrates that the
treatment of source terms proportional to $\nabla \cdot\bB$ is consistent, as discussed in
paper I with reference to \citet{janhunen00} and \citet{dellar01}.

\item The correction terms for a variable smoothing length
may be derived naturally from a variational approach. Accounting for these terms is shown
to improve the accuracy of SPH wave propagation. 
\end{enumerate}

\section*{Acknowledgements} 
DJP acknowledges the support of the Association of Commonwealth Universities and the Cambridge
Commonwealth Trust. He is supported by a Commonwealth Scholarship and Fellowship Plan.

\appendix
\section{Linear waves in MHD}
\label{sec:appendixA}
 In this section we describe the setup used for the MHD waves described in
\S\ref{sec:mwav}. The MHD equations in continuum form may be written as
\begin{eqnarray}
\frac{d\rho}{dt} & = & -\rho \nabla\cdot\bv \label{eq:acty}, \\
\frac{d\bv}{dt} & = & -\frac{\nabla P}{\rho} - \frac{\bB\times(\nabla\times\bB)}{\mu_0\rho},
\label{eq:aind}\\
\frac{d\bB}{dt} & = & (\bB\cdot\nabla)\bv - \bB (\nabla\cdot\bv),
\end{eqnarray}
together with the divergence constraint $\divB = 0$. We perturb according to
\begin{eqnarray}
\rho & = & \rho_0 + \delta\rho, \nonumber \\
\bv & = & \bv, \nonumber \\
\bB & = & \bB_0 + \delta\bB, \nonumber \\
\delta P & = & c_s^2 \delta\rho.
\end{eqnarray}
where $c_s^2 = \gamma P_0 / \rho_0$ is the sound speed. Considering only linear terms, the perturbed equations are therefore given by
\begin{eqnarray}
\frac{d(\delta\rho)}{dt} & = & -\rho_0 (\nabla\cdot\bv), \\
\frac{d\bv}{dt} & = & -c_s^2 \frac{\nabla (\delta\rho)}{\rho_0} - \frac{\bB_0
\times (\nabla\times \delta\bB)}{\mu_0\rho_0}, \\
\frac{d(\delta\bB)}{dt} & = & (\bB_0\cdot\nabla)\bv - \bB_0 (\nabla\cdot\bv).
\end{eqnarray}
Specifying the perturbation according to
\begin{eqnarray}
\delta\rho & = & D e^{i(\bk x - \mathbf{\omega} t)}, \nonumber \\
\bv & = & \bv e^{i(\bk x - \mathbf{\omega} t)}, \nonumber \\
\delta\bB & = & \mathbf{b} e^{i(\bk x - \mathbf{\omega} t)},
\end{eqnarray} 
we have
\begin{eqnarray}
-\omega D & = & -\rho_0 (\bv\cdot\bk) \label{eq:omegaD} \\
-\omega \bv & = & -c_s^2\frac{D\bk}{\rho_0} -
\frac{1}{\mu_0\rho_0}\left[(\bB_0\cdot\mathbf{b})\bk -
(\bB_0\cdot\bk)\mathbf{b}\right] \label{eq:omegav}\\
-\omega \mathbf{b} & = & (\bB_0\cdot\bk)\bv - \bB_0(\bk\cdot\bv)
\label{eq:omegab}.
\end{eqnarray}
Considering only waves in the x-direction (ie. $\bk = [\mathrm{k}_x,0,0]$),
defining the wave speed $v  = \omega/\mathrm{k}$ and using (\ref{eq:omegaD}) to
eliminate $D$, equation (\ref{eq:omegav}) gives
\begin{eqnarray}
v_x \left(v - \frac{c_s^2}{v}\right) & = & \left(\frac{B_{y0} b_y + B_{z0}
b_z}{\mu_0\rho_0}\right), \label{eq:vxeqn} \\
v v_y & = & -\frac{B_{x0} b_y}{\mu_0\rho_0}, \\
v v_z & = & -\frac{B_{x0} b_z}{\mu_0\rho_0},
\end{eqnarray}
where $b_x = 0$ since $\divB = 0$. Using these in (\ref{eq:omegab}) we have
\begin{eqnarray}
v b_y & = & -B_{x0}v_y + B_{y0}v_x, \\
v b_z & = & -B_{x0}v_z + B_{z0}v_x. 
\end{eqnarray}
We can therefore solve for the perturbation amplitudes $v_x,v_y,v_z,b_y$ and $b_z$ in terms of the amplitude
of the density perturbation $D$ and the wave speed $v$. We find
\begin{eqnarray}
v_x & = & \frac{v D}{\rho} \\
v_y \left( v^2 - \frac{B_x^2}{\mu_0\rho} \right) & = & \frac{B_x B_y}{\mu_0\rho} v_x \\
v_z \left( v^2 - \frac{B_x^2}{\mu_0\rho} \right) & = & \frac{B_x B_z}{\mu_0\rho} v_x \\
b_y \left( v^2 - \frac{B_x^2}{\mu_0\rho} \right) & = & v B_y v_x \\
b_z \left( v^2 - \frac{B_x^2}{\mu_0\rho} \right) & = & v B_z v_x 
\end{eqnarray}
 where we have dropped the subscript 0. The wave speed $v$ is found by
eliminating these quantities from (\ref{eq:vxeqn}), giving
\begin{equation}
\frac{v_x}{\left(v^2 - B_x^2/\mu_0\rho \right)}\left[v^4 - v^2\left(c_s^2 +
\frac{B_x^2 + B_y^2 + B_z^2}{\mu_0\rho} \right) + \frac{c_s^2 B_x^2}{\mu_0\rho}\right] = 0,
\label{eq:quartic}
\end{equation}
which reveals the three wave types in MHD. The Alfv\'en waves are those with
\begin{equation}
v^2 = \frac{B_x^2}{\mu_0\rho},
\end{equation}
These are transverse waves which travel along the field
lines. The term in square brackets in (\ref{eq:quartic}) gives a quartic for $v$ (or a quadratic for $v^2$), with roots
\begin{equation}
v^2 = \frac{1}{2}\left[\left(c_s^2 + \frac{B_x^2 + B_y^2 +
B_z^2}{\mu_0\rho}\right) \pm \sqrt{\left(c_s^2 + \frac{B_x^2 + B_y^2 +
B_z^2}{\mu_0\rho}\right)^2 - 4\frac{c_s^2 B_x^2}{\mu_0\rho}} \right],
\end{equation}
which are the fast(+) and slow(-) magnetosonic waves. 

\section{Density perturbation in SPH}
\label{sec:appendixB}
 The perturbation in density is applied by perturbing the particles from an
initially uniform setup. We consider the one dimensional perturbation
\begin{equation}
\rho = \rho_0 [1 + \mathrm{A sin(k}x) ],
\end{equation}
where $A = D/\rho_0$ is the perturbation amplitude. The cumulative total mass in the x
direction is given by
\begin{eqnarray}
M(x) & = & \rho_0\int [1 + \mathrm{A sin (k}x) ] \mathrm{dx} \nonumber \\
& = & \rho_0 [x - \mathrm{ A cos (k}x) ]^{x}_{0},
\end{eqnarray}
such that the cumulative mass at any given point as a fraction of the total mass
is given by
\begin{equation}
\frac{M(x)}{M(x_{max})}.
\end{equation}
For equal mass particles distributed in $x=[0,x_{max}]$ the cumulative mass fraction at particle
$a$ is given by $x_a/x_{max}$ such that the particle position may be calculated
using
\begin{equation}
\frac{x_a}{x_{max}} = \frac{M(x_a)}{M(x_{max})}.
\end{equation}
Substituting the expression for $M(x)$ we have the following equation for the
particle position
\begin{equation}
\frac{x_a}{x_{max}} - \frac{x_a - \mathrm{A cos(k}x_a)}{[x_{max} - \mathrm{A cos(k}x_{max})]} = 0,
\end{equation}
which we solve iteratively using a simple Newton-Raphson rootfinder. With the uniform
particle distribution as the initial conditions this converges in one or two
iterations.

\bibliography{/home/dprice/bibtex/sph,/home/dprice/bibtex/mhd}

\label{lastpage}
\enddocument